\documentstyle[12pt]{article}
\begin{document}
\title{Stochastic modelling of nonlinear dynamical systems}

\author{Piotr Garbaczewski\thanks{Supported by KBN research
grant No 2 P03B 086 16} \\
Institute of Physics, Pedagogical University,\\ 
  pl. S{\l}owia\'{n}ski 6, PL-65 069 Zielona G\'{o}ra,
  Poland}
\maketitle

\begin{abstract}
 We develop a general theory dealing with  stochastic
models for  dynamical systems that are governed by
various nonlinear, ordinary or  partial differential,  equations.
In particular, we  address the problem how
flows in the random medium (related to driving velocity
fields which are generically bound to obey suitable local
conservation laws) can be reconciled with the notion of
 dispersion due to a Markovian diffusion process.
\end{abstract}

\section*{Contexts}

Probabilistic concepts are ubiquitous in diverse areas of nonlinear
science.  Deterministic dynamical systems may give rise to random
transport that is an intrinsic  feature of their  complex behaviour,
augmented by a possible choice of random initial or boundary data
and/or suitable scaling limits.  Nonequlibrium statistical physics
directly employs random processes, Gaussian and non-Gaussian, which
ultimately implement a nonlinear   transport.   Apart from a
concrete identification of genuine sources of randomness  in the
dynamics of classical   systems or in quantum theory,  quite often
we need  quantitative methods  to deal with random-looking
phenomena. Basically that  refers to situations when origins of
randomness are either uncontrollable (allowing merely
for  probabilistic predictions about the future behaviour of the system)
or not  definitely settled.

The so-called Schr\"{o}dinger boundary data and stochastic
interpolation problem  sets a conceptual and formal  basis  for a
surprisingly rich group of topics.
Here, \cite{1}, stochastic analysis
methods are used to deduce the most likely (generally approximate)
underlying dynamics from the given (possibly phenomenological)
input-output statistics data, pertaining to a certain dynamical
 process that is bound to take place in a finite time interval.
The pertinent motion scenarios  range from
 processes arising in   nonequilibrium phenomena, \cite{2,3}, through
classical dynamics of complex systems (deterministic
chaos in terms of densities)
to  searches for a stochastic counterpart of quantum  phenomena.
They  involve random processes that
go beyond the standard Gaussian basis and enable a consistent
usage of the jump-type processes (L\'{e}vy and their perturbed versions)
associated with the anomalous transport, \cite{4}.
In  the diffusion processes  context, we have identified
the third Newton law for
mean velocity fields as being  capable to generate anomalous (enhanced)
or non-dispersive diffusion-type processes through "perturbations
of noise", \cite{3}.

Since  the stochastic interpretation of various
 differential equations is our  major  target, let us mention
 typical  examples that are amenable to our
 methodology. Those are:  Boltzmann, Navier-Stokes, Euler, Burgers
  (more or less standard transport of matter),
Hamilton-Jacobi, Hamilton-Jacobi-Bellmann, Kardar-Parisi-Zhang
(an issue of viscosity solutions and the interface profile growth),
Fokker-Planck, Kramers (standard random propagation related to
the Brownian motion), both linear and nonlinear
Schr\"{o}dinger equations (probabilistic interpretation of solutions,
in   Euclidean and non-Euclidean versions of the problem, also
with reference to an escape over a barrier and decay of a metastable
state).

In most of the above cases a natural linearisation of a nonlinear problem
is provided by generalized diffusion (heat) equations in their
forward and/or backward versions, \cite{1}.
On the contrary, a suitable
coupled pair of time-adjoint nonlinear
diffusion equations  admits a linearisation in terms of the familiar
Schr\"{o}dinger equation. That involves  Markovian diffusion
processes with a feedback (enhancement mechanism named
"the Brownian recoil principle"), \cite{3}.

\section*{Concepts: Schr\"{o}dinger' s interpolation problem}

 There are many procedures to deduce  an  intrinsic dynamics
 of a physical system from observable data. None of them is free of
 arbitrary assumptions needed to reduce the level of ambiguity and
 so none can yield
 a clean choice of the modelling strategy.
 As a standard  example one may invoke the time
 series analysis that is a respected tool in the study
 of complex signals and  is routinely  utilised  for
 a discrimination between  deterministic and  random inputs.

 Our objective is to  reconstruct
 a   \it  random \rm dynamics  that is consistent with
 the given input-output statistics data.
 We shall outline  an algorithm allowing to reproduce  an admissible
   microscopic  motion scenario  under an additional
 assumption that the sought for dynamics actually  \it is \rm
 a   Markovian diffusion process.   This reconstruction method
 is based on solving the so-called Schr\"{o}dinger boundary-data
 and interpolation problem.

 Given two strictly positive (usually on an open space-interval)
 boundary probability densities  $\rho _0(\vec{x}), \rho _T(\vec{x})$
 for a process with the time of duration $T\geq 0$.
One can  single out  a unique Markovian
diffusion process which is specified by solving  the Schr\"{o}dinger
boundary data problem:
\begin{equation}
{m_T(A,B) = \int_A d^3x\int_B d^3y\,  m_T(\vec{x},\vec{y})}
\end{equation}
\begin{equation}
\int d^3y\,  m_T(\vec{x},\vec{y}) = \rho _0(\vec{x}) \, , \,
 \int d^3x\,  m_T(\vec{x},\vec{y})=\rho _T(y)
 \end{equation}
where  the joint probability distribution has a  density
\begin{equation}
{m_T(\vec{x},\vec{y}) = u_0(\vec{x})\, k(\vec{x},0,\vec{y},T)
\, v_T(\vec{y})}
\end{equation}
and the two unknown  functions
$u_0(\vec{x}), \, v_T(\vec{y})$ come out as (unique) solutions, of
\it the same sign, \rm of the integral identities.
To this end, we need to have at our disposal a
continuous bounded  strictly positive (ways to relax this assumption
are known) function
$k(\vec{x},s,\vec{y},t),0\leq s<t\leq T$, which for our purposes
(an obvious
way to secure the Markov property) is chosen to be represented by
familiar Feynman-Kac integral kernels
 of contractive dynamical semigroup operators:
\begin{equation}
{k(\vec{y},s,\vec{x},t)=\int exp[-\int_s^tc(
\vec{\omega }(\tau ),\tau)d\tau ]
d\mu ^{(\vec{y},s)}_{(\vec{x},t)}(\omega )}
\end{equation}

In the above, $d\mu ^{(\vec{y},s)}_{(\vec{x},t)}(\omega)$ is the
conditional
Wiener measure over sample paths of the standard Brownian motion.
(Another choice of the measure allows to extend the framework
to jump-type processes.)

 The pertinent   (interpolating) Markovian
process can be ultimately determined  by means of
positive solutions (it is desirable to have them bounded)
 of the adjoint pair of  generalised heat equations:
\begin{equation}
{\partial _tu(\vec{x},t)=\nu \triangle u(\vec{x},t) -
c(\vec{x},t)u(\vec{x},t)}
\end{equation}
\begin{equation}
\partial _tv(\vec{x},t)= -\nu \triangle v(\vec{x},t) +
c(\vec{x},t)v(\vec{x},t)\enspace .
\end{equation}

Here, a function $c(\vec{x},t)$
is restricted only by the positivity and continuity demand
for the kernel.

Solutions, upon suitable normalisation give rise to the
Markovian  diffusion process with the \it factorised \rm
probability density $\rho (\vec{x},t)=u(\vec{x},t)v(\vec{x},t)$
which, while evolving in time, interpolates between
the boundary density data $\rho (\vec{x},0)$ and 
$\rho (\vec{x},T)$. The interpolation admits an It\^{o} realisation
  with the respective forward and
backward drifts  defined as follows:
\begin{equation}
{\vec{b}(\vec{x},t)=2\nu {{\nabla v(\vec{x},t)}
\over {v(\vec{x},t)}}}
\end{equation}
\begin{equation}
\vec{b}_*(\vec{x},t)= - 2\nu {{\nabla u(\vec{x},t)}
\over {u(\vec{x},t)}}
\end{equation}
in the prescribed time interval $[0,T]$.\\
For the forward interpolation, the familiar Fokker-Planck
(second Kolmogorov) equation holds true:
\begin{equation}
{\partial _t\rho (\vec{x},t) = \nu \triangle
\rho (\vec{x},t) - \nabla [\vec{b}(\vec{x},t)\rho (\vec{x},t)]}
\end{equation}
with $\rho (\vec{x},0)$ given, while for the backward
interpolation (starting from $\rho (\vec{x},T)$) we have:
\begin{equation}
{\partial _t\rho(\vec{x},t) = - \nu \triangle \rho (\vec{x},t) -
\nabla [\vec{b}_*(\vec{x},t) \rho (\vec{x},t)]\enspace . }
\end{equation}

The drifts are
gradient fields, $curl \, \vec{b}= 0$. As a consequence,
those  that are allowed by any  prescribed choice of   the function
$c(\vec{x},t)$  \it must \rm fulfill the compatibility
condition
\begin{equation}
{c(\vec{x},t) = \partial _t \Phi \, +\,
{1\over 2} ({b^2
\over {2\nu }}+ \nabla b)}
\end{equation}
which establishes the Girsanov-type  connection of
the forward drift
$\vec{b}(\vec{x},t)=2\nu \nabla \Phi (\vec{x},t)$ with the
Feynman-Kac potential  $c(\vec{x},t)$.
In the considered Schr\"{o}dinger's interpolation
framework, the forward and backward
drift fields  are  connected   by the identity
$\vec{b}_*= \vec{b} - 2\nu \nabla ln \rho $.

For  Markovian diffusion processes the
notion of the \it backward \rm transition probability density
 $p_*(\vec{y},s,\vec{x},t)$ can be consistently introduced on 
each finite  time interval, say $0\leq s<t\leq T$:
 \begin{equation}
 {\rho (\vec{x},t) p_*(\vec{y},s,\vec{x},t)=
 p(\vec{y},s,\vec{x},t) \rho (\vec{y},s)} 
\end{equation}
so that $\int \rho (\vec{y},s)p(\vec{y},s,\vec{x},t)d^3y=
\rho (\vec{x},t)$
and $\rho (\vec{y},s)=\int p_*(\vec{y},s,\vec{x},t)
\rho (\vec{x},t)d^3x$.  

The transport (density evolution) equations  refer to
processes running
in opposite  directions  in a fixed, common for both
time-duration period.
The  forward one executes an interpolation from the Borel set $A$
to $B$, while the  backward one executes  an interpolation from
$B$ to $A$.

Let us mention at this point that
various partial differential equations associated with Markovian
diffusion processes are known \it not \rm to be invariant under
time reversal. That implies dissipation and links them with  irreversible
physical phenomena.
However,  the correspoding processes are known to admit a \it statistical
inversion   \rm    and asking for a statistical past of the process makes
sense.

      In particular,
the knowledge of the Feynman-Kac kernel  implies that the
transition probability density of the forward  process reads:
\begin{equation}
{p(\vec{y},s,\vec{x},t)=k(\vec{y},s,\vec{x},t)
{{v(\vec{x},t)}\over {v(\vec{y},s)}}\enspace .}
\end{equation}
while the corresponding
 transition
probability density  of the backward process has the form:
\begin{equation}
{p_*(\vec{y},s,\vec{x},t) = k(\vec{y},s,\vec{x},t)
{{u(\vec{y},s)}\over {u(\vec{x},t)}}\enspace .}
\end{equation}

Obviously in the time interval $0\leq s<t\leq T$
there holds:
\begin{equation}
{u(\vec{x},t)=\int u_0(\vec{y}) k(\vec{y},s,\vec{x},t) d^3y}
\end{equation}
\begin{equation}
v(\vec{y},s)=\int k(\vec{y},s,\vec{x},T) v_T(\vec{x})d^3x
\enspace .
\end{equation}

Consequently, we have fully determined the underlying  (Markovian)
random motions, forward and backward, respectively.
All that accounts for perturbations of (and conditioning upon)
the Wiener noise. \\

\section*{Particularities}

If we consider  a   fluid in thermal equilibrium as the noise
 carrier, a kinetic theory viewpoint  amounts to visualizing the
 constituent molecules that collide
not only with each other but  also  with the tagged (colloidal)
particle, so \it enforcing \rm  and \it maintaining  \rm
its  observed erratic  motion.
The Smoluchowski  approximation  takes  us   away
 from those kinetic theory intuitions by  projecting  the
phase-space theory of random motions into its  configuration
space image  which is a spatial Markovian diffusion
process, whose formal infinitesimal encoding reads:
\begin{equation}
{d\vec{X}(t)= {\vec{F}\over {m\beta }}dt +
\sqrt{2D}d\vec{W}(t)\enspace .}
\end{equation}

In the above $m$ stands for the mass of a diffusing particle,
$\beta $ is a friction parameter, D is a diffusion constant
 and $\vec{W}(t)$ is a normalised Wiener process.
The Smoluchowski forward drift can be traced back to
a presumed selective action of the external force $\vec{F}=
-\vec{\nabla }V$ on the Brownian particle
that  has a negligible  effect on the thermal bath but
in view of frictional  resistance imparts to a particle
the \it  mean \rm velocity $\vec{F}/m\beta $ on
the $\beta ^{-1}$ time scale.

The
noise carrier (fluid in the present considerations)
statistically remains in the state of rest, with \it no \rm
intrinsic mean flows, hence  unperturbed in the mean (all other
cases we associate with the term "perturbations of noise").
At the first glance, a formal
replacement of the Smoluchowski forward drift in Eq. (17)
by any   space-time dependent driving velocity  field
would suggest  a legitimate  procedure
to implement   a net  transport that would   combine
dispersion due to a diffusion process with deterministic
mean flows due to external agencies
(consider Euler, Navier-Stokes or   Burgers velocity
fields as reference examples).
However the situation is not that simple.

It is well known that
a spatial diffusion (Smoluchowski) approximation of the
phase-space process, allows
to reduce the number of \it independent \rm
local conservation laws to two only.
Therefore the Fokker-Planck  equation can
always be supplemented  by another (independent) partial
differential equation to form a \it closed \rm system.
 If we assign
 a probability density $\rho _0(\vec{x})$ with which the
 initial data
  $\vec{x}_0=\vec{X}(0)$  are
distributed, then the emergent Fick law  would  reveal a
statistical tendency of particles to flow away from
higher probability  residence areas.     This
feature is encoded in the corresponding Fokker-Planck equation:
${\partial _t \rho  = - \vec{\nabla }\cdot (\vec{v}\rho )=
- \vec{\nabla }\cdot [({\vec{F}\over {m\beta }}  - D
{{\vec{\nabla }\rho }\over {\rho }}) \rho ]}$
where a diffusion current velocity  is
$\vec{v}(\vec{x},t) = \vec {b}(\vec{x},t) - D{{\vec{\nabla }\rho
(\vec{x},t)}\over {\rho (\vec{x},t)}}$
while  the forward drift reads $\vec{b}(\vec{x},t) =
{\vec{F}\over {m\beta }}$.
Clearly, the local diffusion current (a local flow that might
be experimentally observed for  a cloud of suspended particles
in a liquid)
$\vec{j}=\vec{v} \rho $
gives rise to a non-negligible  matter transport  on the
ensemble average, even if no intrinsic mean flows are in
existence in the random medium.

We may surely consider a formal replacement of  the previous
${\vec{F}\over {m\beta }}$  by a local velocity field
 $\vec{v}(\vec{x},t)$. However, irrespectively  of whether
 we utilize ${\vec{F}\over {m\beta }}$  or   $\vec{v}(\vec{x},t)$
in the formalism, the  velocity field \it must   \rm
  obey the natural (local) momentum  conservation law
which   directly originates from
the rules of the It\^{o} calculus for Markovian diffusion processes
 and from the first moment equation in the
diffusion approximation  (!) of the Kramers theory:
\begin{equation}
{\partial _t\vec{v} + (\vec{v} \cdot \vec{\nabla }) \vec{v} =
\vec{\nabla }(\Omega - Q)\enspace .}
\end{equation}

An effective  potential function $\Omega (\vec{x})$
can be expressed in terms of the  Smoluchowski forward drift
$\vec{b}(\vec{x}) = {\vec{F}(\vec{x})
\over {m\beta }}$ as follows:
$\Omega = {{\vec{F}^2} \over {2m^2\beta ^2}} + {D\over {m\beta }}
\vec{\nabla } \cdot \vec{F}$. That is to be compared with
Eq. (11) which governs the more general space-time dependent
situation.

Moreover, in the would-be Euler equation (18), instead of the
standard pressure term,
 there appears a contribution from a
  probability density  $\rho $-dependent potential
 $Q(\vec{x},t)$. It  is given in terms of the so-called osmotic
velocity field  $\vec{u}(\vec{x},t) =
D\vec{\nabla } \, ln \rho (\vec{x},t)$:
$Q(\vec{x},t) = {1\over 2} \vec{u}^2 + D\vec{\nabla }
\cdot \vec{u}$
and  is  generic to a local momentum conservation
law  respected by   isothermal Markovian diffusion processes.

To  analyze  general  perturbations  of the
medium and then the resulting intrinsic (mean) flows,
a  function
$\vec{b}(\vec{X}(t),t)$,  must  replace the
Smoluchowski drift.
Under suitable restrictions, we can relate probability  measures
corresponding to  different (in terms of forward drifts !)
Fokker-Planck equations  and  processes by means of the
Cameron-Martin-Girsanov theory  of measure transformations,
\cite{1}.
For suitable  forward drifts which
are gradient fields   that  yields
the most  general  form of an auxiliary potential  (cf. Eq. (11))
${\Omega (\vec{x},t) = 2D[ \partial _t\phi + {1\over 2}
({\vec{b}^2\over {2D}} + \vec{\nabla }\cdot \vec{b})]\enspace .}$
We denote $\vec{b}(\vec{x},t) = 2D \vec{\nabla } \phi (\vec{x},t)$.

   Mathematical features of the formalism
appear to depend crucially on the properties (like continuity,
local and global boundedness, Rellich class) of the
auxiliary potential $\Omega $.
Let us consider
a bounded from below (local boundedness from above is useful as well),
 continuous function $\Omega (\vec{x},t)$.
 Then,  by means of the  gradient
field ansatz for the diffusion current velocity
($\vec{v}=\vec{\nabla }S\rightarrow
\partial _t\rho = - \vec{\nabla }\cdot [(\vec{\nabla } S)\rho ]$)
we can  transform the local momentum conservation law of a
Markovian diffusion process to  a universal
Hamilton-Jacobi form:
\begin{equation}
{\Omega = \partial _tS + {1\over 2} |\vec{\nabla }S|^2  + Q }
\end{equation}
where $Q(\vec{x},t)$ was defined before. By
applying the  gradient operation  we recover the conservation law
(18).

In the above,  the contribution due to $Q$
is a direct consequence of  an initial probability measure choice
for the diffusion    process,
 while $\Omega $  does account for an appropriate
forward drift of the process via  Eq. (19).

Thus, in the context of Markovian diffusion processes,
 we can consider a  \it closed \rm system of partial
 differential equations
 which comprises   the continuity
equation $\partial _t \rho =- \vec{\nabla }(\vec{v}\rho )$
 and  the Hamilton-Jacobi  equation, plus
 suitable initial (and/or boundary) data.
The underlying isothermal diffusion process is specified uniquely.
Those two partial differential equations set ultimate limitations
in the old-fashioned  problem of "how much nonlinear" and
"how much space-time dependent"  the driving velocity field can be
to yield a  consistent stochastic diffusion process (or the
Langevin-type dynamics)

\end{document}